\documentclass[aps,prr,reprint,twocolumn,longbibliography,superscriptaddress,draft=false]{revtex4-1}

\usepackage{amsmath, amsfonts, bm, bbm, braket, mathtools, comment, slashed}
\usepackage[inline]{enumitem}
\usepackage{hyperref}
\hypersetup{
  pdftex,
  pdftitle={Insights into the Anisotropic Spin-S Kitaev Chain},
  pdfauthor={Jacob S. Gordon and Hae-Young Kee},
  colorlinks=true,
  linkcolor=blue,
  citecolor=blue,
  filecolor=blue,      
  urlcolor=blue,
  allcolors=blue
} 
          
\usepackage{natbib}
\usepackage{graphicx}
\usepackage{siunitx}
\usepackage{multirow}
\usepackage{lipsum}

\newcommand{\pmat}[1]{\begin{pmatrix} #1 \end{pmatrix}} % matrices
\newcommand{\id}{\mathbbm{1}}							% identity operator

\DeclareMathOperator{\sgn}{\mathrm{sgn}}

\newcommand{\Z}{\mathbb{Z}}
\newcommand{\C}{\mathcal{C}}
\newcommand{\D}{\mathcal{D}}

\renewcommand{\H}{\mathcal{H}}

\renewcommand{\P}{\mathcal{P}}
\newcommand{\Q}{\mathcal{Q}}
\newcommand{\T}{\mathcal{T}}
\newcommand{\Tds}{\mathbb{T}}
\newcommand{\J}{\mathcal{J}}

\newcommand{\xh}{\bm{\hat{x}}}
\newcommand{\yh}{\bm{\hat{y}}}
\newcommand{\zh}{\bm{\hat{z}}}

\renewcommand{\a}{\bm{\hat{a}}}
\renewcommand{\b}{\bm{\hat{b}}}
\renewcommand{\c}{\bm{\hat{c}^*}}

\renewcommand{\S}{\bm{S}}

\newcommand{\vp}{\varphi}

\renewcommand{\a}{\hat{a}^{\phantom{\dag}}}
\newcommand{\ad}{\hat{a}^{\dag}}
\renewcommand{\b}{\hat{b}^{\phantom{\dag}}}
\newcommand{\bd}{\hat{b}^{\dag}}
\newcommand{\g}{\hat{\gamma}^{\phantom{\dag}}}
\newcommand{\gd}{\hat{\gamma}^{\dag}}

\makeatletter
\g@addto@macro\bfseries{\boldmath}
\makeatother

\begin{document}

\title{Insights into the Anisotropic Spin-$S$ Kitaev Chain}

\author{Jacob S. Gordon}
\affiliation{Department of Physics, University of Toronto, Ontario M5S 1A7, Canada}
\author{Hae-Young Kee}
\email{hykee@physics.utoronto.ca}
\affiliation{Department of Physics, University of Toronto, Ontario M5S 1A7, Canada}
\affiliation{Canadian Institute for Advanced Research, CIFAR Program in Quantum Materials, Toronto, ON M5G 1M1, Canada}

\date{\today}

\begin{abstract}
  Recently there has been a renewed interest in properties of the higher-spin Kitaev models, especially their low-dimensional analogues with additional interactions.
  These quasi-1D systems exhibit rich phase diagrams with symmetry-protected topological phases, Luttinger liquids, hidden order, and higher-rank magnetism.
  However, the nature of the pure spin-$S$ Kitaev chains are not yet fully understood.
  Earlier works found a unique ground state with short-ranged correlations for $S = 1$, and an intriguing double-peak structure in the heat capacity associated with an entropy plateau. 
  To understand the low-energy excitations and thermodynamics for general $S$ we study the anisotropic spin-$S$ Kitaev chain.
  Starting from the dimerized limit we derive an effective low-energy Hamiltonian at finite anisotropy.
  For half-integer spins we find a trivial effective model, reflecting a non-local symmetry protecting the degeneracy, while for integer $S$ we find interactions among the flux degrees of
  freedom that select a unique ground state.
  The effective model for integer spins is used to predict the low-energy excitations and thermodynamics, and we make a comparison with the semiclassical limit through linear spin wave theory.
  Finally, we speculate on the nature of the isotropic limit.
\end{abstract}

\maketitle

\section{Introduction}

The integrable Kitaev honeycomb model~\cite{kitaev2006honeycomb} and its generalizations have been a major topic of study in modern condensed matter physics.
For $S = \tfrac{1}{2}$ the honeycomb model hosts fractionalized excitations, including free Majorana fermions on a static $\Z_2$ gauge field.
Large bond anisotropy drives the system into a gapped state equivalent to the toric code in the low-energy limit, and generic magnetic fields impart an
effective topological term which leads the system into an Ising topological order phase.
These non-trivial phases have garnered attention for their potential in quantum computing.

Initially an academic problem, the potential of the $S = \tfrac{1}{2}$ Kitaev model for describing real honeycomb $d^5$ Mott insulators was discovered by~\citet{jk2009prl} through
an interplay between electron correlations and spin-orbit coupling.
Models with additional symmetry-allowed exchange interactions were subsequently put forward~\cite{rau2014prl}, and found
relevance~\cite{jk2009prl, rau2014prl, cjk2010prl, singh2012relevance, plumb2014prb, witczakkrempa2014correlated,rau2016novel,trebst2017kitaev,winter2017models,motome2020design,takayama2021phases}
in describing the candidates A$_2$IrO$_3$~\cite{jk2009prl,rau2014prl,cjk2010prl,singh2012relevance} and $\alpha$-RuCl$_3$~\cite{plumb2014prb}.
Recently there has been a renewed interest in the higher-spin analogues of the honeycomb model, including concrete proposals for their realization.
\citet{stavropoulos2019higher} have suggested that Hund's coupling on transition metal sites can stabilize a larger moment, while significant spin-orbit
coupling on the surrounding ligands can produce bond-dependent interactions.
Despite not being exactly solvable, the higher-spin analogues have many features in common with the $S = \tfrac{1}{2}$ model~\cite{baskaran2008spinS}, such as an extensive number of conserved
quantities, short-range spin correlations, thermodynamics~\cite{koga2018thermodynamic,oitmaa2018incipient}, and behaviour under magnetic field~\cite{zheng2020liquid,hickey2020spin,khait2021character}.
However, there also seem to be some important differences, including the topological nature of the isotropic limit~\cite{rousochatzakis2018regime,lee2020tensor}
and the effective description with anisotropy~\cite{minakawa2019behaviour,lee2021anisotropy}.

In addition to variations in spin length, there has been an explosion of research into lower-dimensional analogues of the
Kitaev model~\cite{feng2007characterization,brzezicki2007compass,you2008pseudospin}
with additional interactions~\cite{steinigeweg2016energy,agrapidis2018ordered,vimal2018correlations,catuneanu2019nonlocal,agrapidis2019ladder,yang2020emergentSU2,yang2020comprehensive,yang2020spinwave,liu2020lifshitzG,yang2021KGchain,luo2021unveiling},
and/or external magnetic fields~\cite{sun2009field,subrahmanyam2013block,gordon2019theory,sorensen2021entanglement,metavitsiadis2021flux}.
The study of these quasi-1D chains and ladders with additional interactions has revealed a plethora of interesting phases, including symmetry-protected topological phases (SPTs), disordered phases with hidden order, Luttinger liquids, and higher-rank magnetism.
An early work by~\citet{sen2010spin1kitaev} showed that the pure spin-$S$ Kitaev chains have an analogue of the $\Z_2$-valued conserved quantities, and that there is a qualitative
difference between the integer and half-integer spin due to their properties.
They show that the flux-free sector for any integer $S$ has fractional quantum dimension, and that the $S = 1$ chain hosts a unique ground state with local excitations
of the $\Z_2$ conserved quantities.
This picture for the $S = 1$ chain was recently confirmed by~\citet{luo2021unusual}, where they discovered a double-peak structure in the heat capacity, and a singly-degenerate
entanglement spectrum that rules out an SPT.

Motivated by these results, we have focused on the anisotropic spin-$S$ Kitaev chain and it's classical analogue to gain further insight into these properties and
the expected behaviour for general $S$.
After reviewing the model in Sec.~\ref{sec:conserved}, in Sec.~\ref{sec:perturbation} we find an effective description of the low-energy physics in the anisotropic limit of the spin-$S$ chain.
The effective Hamiltonian for integer spins emphatically selects the flux-free sector to all orders of perturbation theory, while the protected degeneracy in the
half-integer spins manifests as a trivial effective model.
This structure in the integer spin effective model naturally explains the ground state with periodic boundary conditions (PBCs) and open boundary conditions (OBCs), excitations, and expected
low-energy thermodynamics, which we compare across anisotropies and spin length.
Next, we review the classical ground states of the Kitaev chain in Sec.~\ref{sec:semiclassical} and the degeneracy when spin wave fluctuations are taken into account.
In the limit of $S \rightarrow \infty$ the two pictures coalesce, and we regain the degeneracy predicted by linear spin wave theory.

\section{Model \& Conserved Quantities}\label{sec:conserved}

\begin{figure}
  \includegraphics[width=0.75\columnwidth]{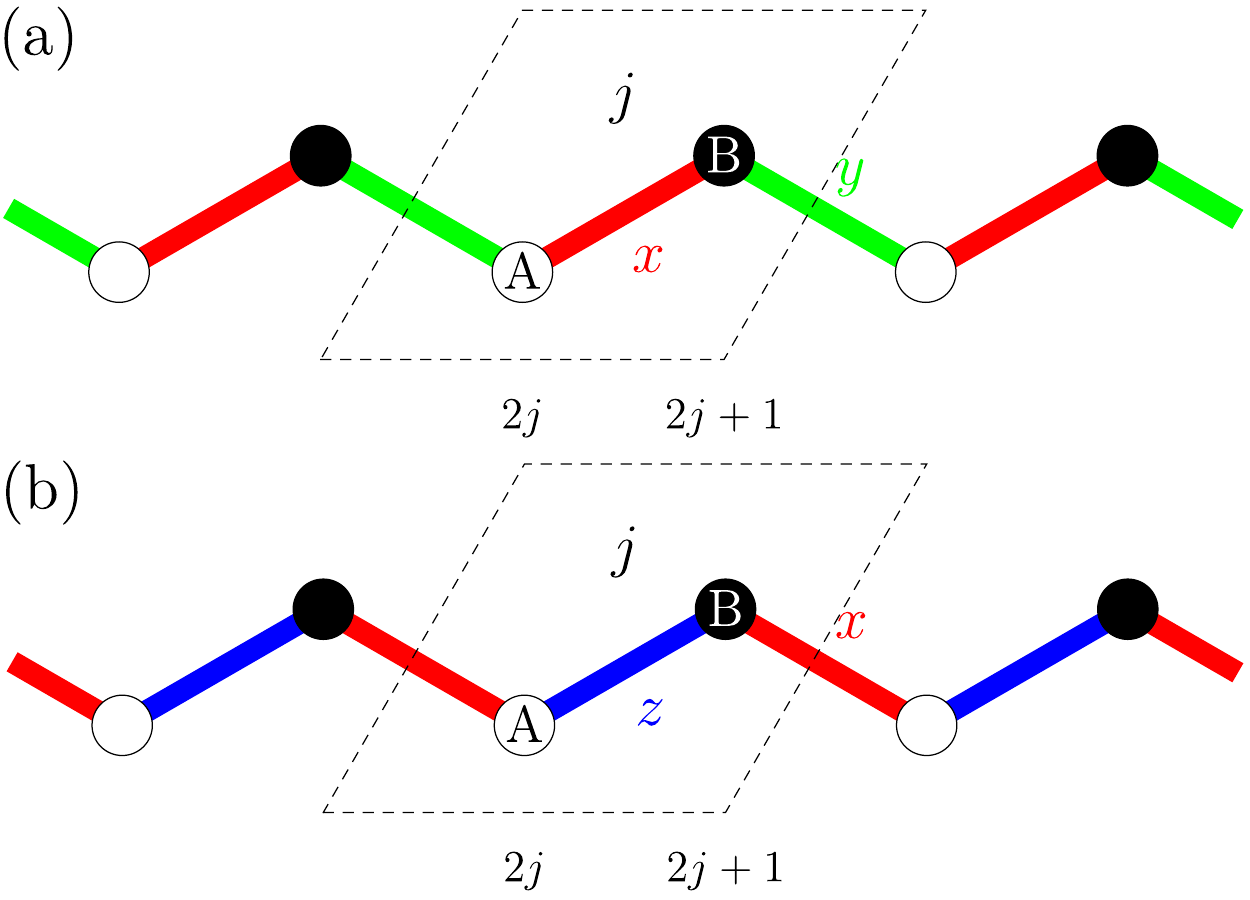}
  \caption{
    Definition of the site and unit cell labels for the one-dimensional Kitaev model.
    The unit cell, indexed by $j = 0,1,\cdots,N_c-1$, contains the A and B sublattices at even and odd integers, respectively.
    (a) Usual definition with the $x$- and $y$-bonds along the chain, and (b) a $\C_{3\c}$ spin-rotated version for discussing
    the perturbation theory in the anisotropic limit.
  }
  \label{fig:chains}
\end{figure}  

In this section we discuss the one-dimensional analogue of the spin-$S$ Kitaev model.
The Hamiltonian takes the usual form
\begin{equation}\label{eq:kitaevchainH}
  \H = \sum_{j = 0}^{N_c -1}\left(K_0 S_{j,A}^xS_{j,B}^x + K_1 S_{j,B}^yS_{j+1,A}^y\right),
\end{equation}
where $j$ indexes the $N_c$ unit cells as depicted in Fig.~\ref{fig:chains}(a).
We impose PBCs on the $N_s = 2N_c$ sites through $S_{j + N_c,A/B}^{\alpha} = S_{j,A/B}^{\alpha}$.
OBCs can be imposed by omitting the second term with $j = N_c-1$.

As in the spin-$S$ honeycomb model~\cite{baskaran2008spinS}, the Kitaev chain has an extensive number of conserved quantities~\cite{sen2010spin1kitaev}.
The conserved quantities are built from the $\pi$-rotation operators about the Cartesian spin axes
\begin{equation}
  \Sigma^{\alpha} = e^{-i\pi S^{\alpha}}.
\end{equation}
These operators have the property that $(\Sigma^{\alpha})^2 = e^{-2\pi i S^{\alpha}} = (-1)^{2S}$, implying that the eigenvalues are
$\pm 1$ for integer $S$ and $\pm i$ for half-integer $S$.
Another consequence is $(\Sigma^{\alpha})^{\dag} = (-1)^{2S}\Sigma^{\alpha}$, from which we can derive
\begin{equation}
  \Sigma^{\alpha}\Sigma^{\beta} - (-1)^{2S}\Sigma^{\beta}\Sigma^{\alpha} = 0 \qquad \alpha\ne\beta.
\end{equation}
Therefore, the $\Sigma^{\alpha}$ commute for integer $S$, and anti-commute for half-integer $S$.
On each of the bond types we can define the quantities
\begin{equation}
  W_{j,0} = \Sigma_{j,A}^y\Sigma_{j,B}^y, \qquad W_{j,1} = \Sigma_{j,B}^x\Sigma_{j+1,A}^x.
\end{equation}
Both operators square to $(-1)^{4S} = +\id$ (and thus have $\Z_2$ eigenvalues $\pm 1$ for any $S$), and commute with the Hamiltonian.
In the integer spin case $[W_{j,0},W_{k,1}] = 0$ since the $\Sigma_k^{\alpha}$ commute with each other, so we can simultaneously diagonalize
them with the Hamiltonian.
For half-integer $S$ we cannot simultaneously diagonalize them due to the anti-commutation of neighbouring $W_{j,0}$ and $W_{k,1}$.
As pointed out by \citet{sen2010spin1kitaev} this property has a special implication best seen through the non-local operators
\begin{equation}\label{eq:nonlocalSU2}
  \mu_j^x = \prod_{k \le j}W_{k,0}, \qquad \mu_j^z = W_{j,1}, \qquad \mu_j^y = i\mu_j^x\mu_j^z.
\end{equation}
These commute with the Hamiltonian, and satisfy $N_c$ independent SU(2) algebras
\begin{equation}
  [\mu_j^{\alpha},\mu_k^{\beta}] = 2i \delta_{jk}\epsilon^{\alpha\beta\gamma}\mu_j^{\gamma}, \qquad
  \{\mu_j^{\alpha},\mu_j^{\beta}\} = 2\delta^{\alpha\beta}\id.
\end{equation}
This implies a degeneracy of (at least) $2^{N_c} = 2^{N_s/2}$ for each eigenstate in the half-integer spin model.

\section{Perturbation Theory in the Anisotropic Limit}\label{sec:perturbation}

In this section we consider the low-energy description of the general spin-$S$ anisotropic Kitaev chain.
Without loss of generality we consider $|K_0| \gg |K_1|$, as the other limit works out similarly.
For the purposes of the perturbation theory we perform a global $\C_{3\c}$ spin rotation to diagonalize the strong $K_0$ bonds as
shown in Fig.~\ref{fig:chains}(b)
\begin{equation}
  \H_0 = K_0\sum_{j = 0}^{N_c-1}S_{j,A}^zS_{j,B}^z, \quad \H_1 = K_1\sum_{j = 0}^{N_c-1}S_{j,B}^xS_{j+1,A}^x.
\end{equation}
The conserved quantities are also rotated into
\begin{equation}
  W_{j,0} = \Sigma_{j,A}^x\Sigma_{j,B}^x, \qquad W_{j,1} = \Sigma_{j,B}^z\Sigma_{j+1,A}^z.
\end{equation}
In the extreme anisotropic limit $K_1 = 0$ the Hamiltonian consists of $N_c$ disconnected $z$-dimers which are diagonal in the global
$S^z$ basis $(\ket{m_{j,A}}\otimes\ket{m_{j,B}})^{\otimes N_c}$ with $m\in\{S,S-1,\cdots,-S\}$ for a degeneracy of $2^{N_c} = 2^{N_s/2}$.
Using the fact that the time-reversal operator acts via $\T\ket{m} = (-1)^{S-m}\ket{-m}$, we choose time-reversal partners as a basis for
the $z$-dimers
\begin{equation}
  \{\ket{a_j},\ket{b_j}\} = 
  \begin{cases}
    \{\ket{+S,-S},(-1)^{2S}\ket{-S,+S}\} & K_0 > 0 \\
    \{\ket{+S,+S},\ket{-S,-S}\} & K_0 < 0
  \end{cases}.
\end{equation}
This allows us to define a set of Pauli matrices for the low-energy subspace of the dimer via
\begin{align}
\begin{split}  
  \tau_j^x &= \ket{a_j}\bra{b_j} + \ket{b_j}\bra{a_j}, \\
  \tau_j^y &= -i\ket{a_j}\bra{b_j} +i \ket{b_j}\bra{a_j}, \\
  \tau_j^z &= \ket{a_j}\bra{a_j} - \ket{b_j}\bra{b_j}.  
\end{split}  
\end{align}  
With the relations $\Sigma^z\ket{m} = (-1)^m\ket{m}$ and $\Sigma^x\ket{m} = (-1)^S\ket{-m}$ we can work out expressions for the $W_{j,0/1}$ in
the low-energy subspace
\begin{equation}\label{eq:Wlowenergy}
    W_{j,0} \mapsto (\sgn K_0)^{2S}\tau_j^x,\quad    W_{j,1} \mapsto (\sgn K_0 \tau_j^z\tau_{j+1}^z)^{2S}.
\end{equation}
Our goal is now to obtain the effective Hamiltonian $\H_{\text{eff}} = \sum_{n = 0}^{\infty}\H_{\text{eff}}^{(n)}$ containing the terms
\begin{equation}\label{eq:effectiveH}
   \H_{\text{eff}}^{(n)} = \begin{cases}
     						\P\H_0\P = E_0\P & n = 0 \\
							\P\left[\H_1\Q\frac{1}{E_0 - \H_0}\Q\right]^{n-1}\H_1\P & n > 0
						 \end{cases}                            
\end{equation}
where $\P$ projects onto the low-energy subspace, and $\Q = \id - \P$.
The structure of the problem allows us to identify all possible terms generated in this expansion.
To clearly see the reason why we will discuss the perturbative processes on clusters of increasing size, noting the conditions for a non-zero contribution.

\begin{figure}
  \includegraphics[width=1.01\linewidth]{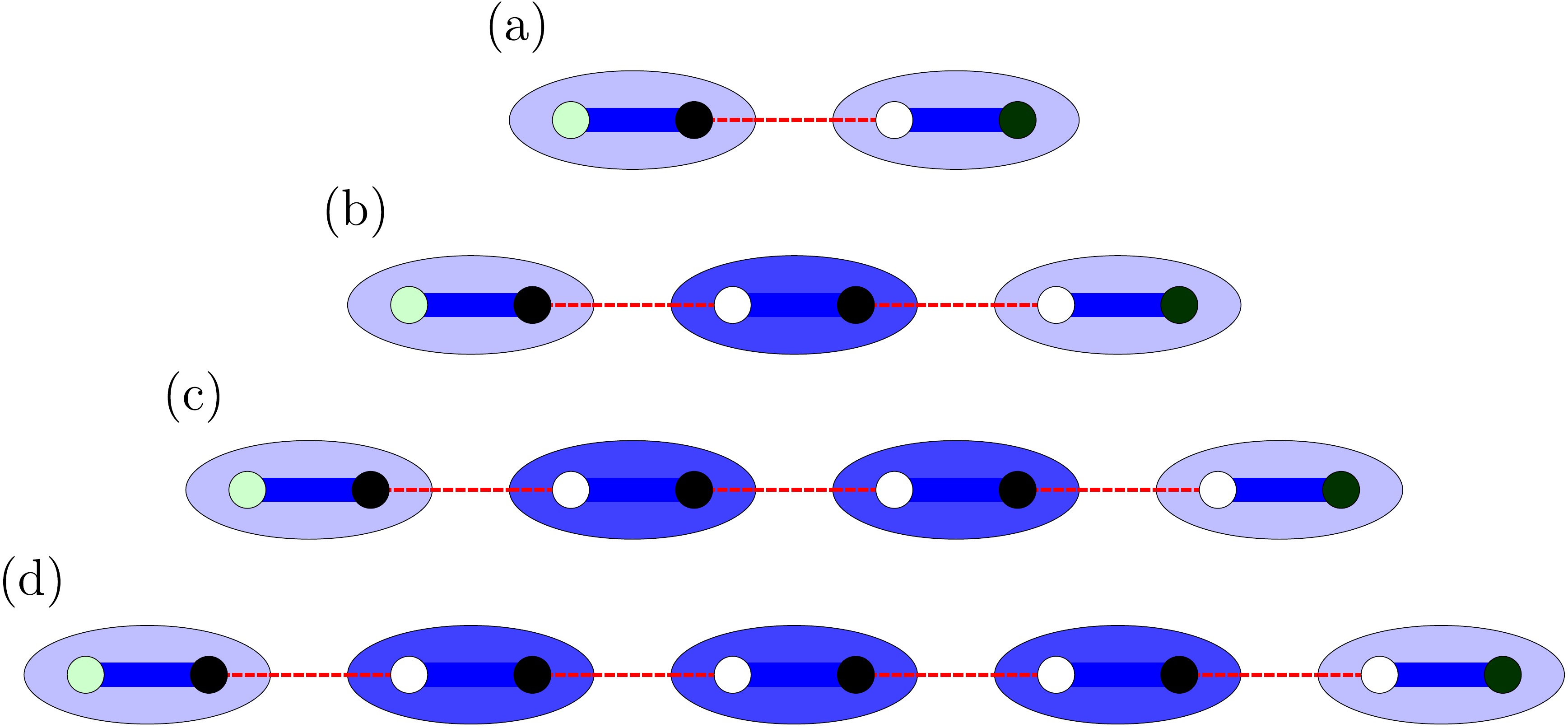}
  \caption{
    Clusters of $z$-dimers connected by $x$-bonds in $\H_1$  used to derive terms in the effective Hamiltonian.
    Edge dimers are shaded more lightly than the inner dimers, and the outermost spins not affected by the $\H_1$ bonds are tinted green.
    The two dimer cluster in (a) can only generate trivial terms, as their dimer partners cannot change their spin.
    To flip the pseudospin of the inner dimer in (b), the two inner dimers in (c), or the three inner dimers in (d), $2S$ applications of $\H_1$ on each bond are required.
    In order for the edge dimers to return to the ground state manifold, the number of $\H_1$ applications on the outer bonds must be even.
  }
  \label{fig:dimers}
\end{figure}  

The simplest problem is that of two dimers connected by a single bond, as shown in Fig.~\ref{fig:dimers}(a).
Our perturbation is $S^xS^x$ containing $S^{\pm}S^{\pm} + S^{\pm}S^{\mp}$, which raises or lowers the inner spins.
Note that the outermost spins, tinted green in Fig.~\ref{fig:dimers},  are not changed by this operator.
At the end of the perturbative processes represented in Eq.~(\ref{eq:effectiveH}) we must return to one of the ground states to survive the projection $\P$.
Since the outermost spins are fixed the inner spins they form dimers with must also return to their initial state.
For this to happen we require that $\H_1$ has been applied on the $x$-bond an {\it even} number of times so as to undo every spin raising/lowering that took place.
As each dimer is forced to return to its initial state we generate an identity term in the low-energy subspace.

Our next rule comes from considering a cluster of three dimers with two $\H_1$ bonds as shown in Fig.~\ref{fig:dimers}(b).
Again the outermost spins are not changed by the inner $\H_1$ bonds, so the spins they form dimers with must also return to their initial state.
This means that we must apply $\H_1$ on each $x$-bond an even number of times.
What are the possibilities for the inner dimer?
Both spins of the inner dimer are now free to change during the perturbative process, but as always must return to a member of the ground state manifold.
This leaves us with two possibilities.
Either both spins of the inner dimer return to their initial state, e.g. $\ket{S,S} \mapsto \ket{S,S}$, or both flip, e.g. $\ket{S,S} \mapsto \ket{-S,-S}$.
The first scenario is always possible and represents an identity term in the low-energy subspace.
If possible, the second scenario is an effective spin flip of the inner dimer represented by $\tau^x$.
For a spin $S$ state to completely flip $\ket{S} \mapsto \ket{-S}$ we require $2S$ applications of $S^x$ operators involving that site.
In order to have both spins of the inner dimer flip, we require $2S$ applications of $S^xS^x$ on each protruding bond for a total of $2S + 2S = 4S$.

With both these rules we are in a position to enumerate the possible terms, and see the difference between integer and half-integer spins.
Staying with the three dimer cluster of Fig.~\ref{fig:dimers}(b), we can ask if the process that flips the inner dimer pseudospin is compatible with our
requirement that an {\it even} number of $S^x$ terms be applied to the inner spin of the edge dimers.
To completely flip a spin in the inner dimer we would need to apply $S^xS^x$ $2S$ times on the outer bonds in the process, which is an {\it odd} number
for half-integer spins.
This odd number of applications does not allow the inner spin in the edge dimer to return back to its initial state, and is killed by the final projection $\P$.
On the other hand, $2S$ is even for integer spins which allows the outer dimer to return to its initial state.

\subsection{Half-Integer Spins}

To see these rules in action, let's consider a process for $S = \tfrac{1}{2}$ on three dimers for $K_0 < 0$.
Both inner spins of the central dimer can be flipped by applying $S^xS^x$ on each bond $2S = 1$ time
\begin{align}
  \begin{split}    
    &\ket{\uparrow\uparrow}{\color{red}-}\ket{\uparrow\uparrow}{\color{red}-}\ket{\uparrow\uparrow} \\
    &\hspace{2.5mm}S^-\downarrow S^- \\
    &\ket{\uparrow\downarrow}{\color{red}-}\ket{\downarrow\uparrow}{\color{red}-}\ket{\uparrow\uparrow} \\
    &\hspace{12mm}S^-\downarrow S^- \\
    &\ket{\uparrow\downarrow}{\color{red}-}\ket{\downarrow\downarrow}{\color{red}-}\ket{\downarrow\uparrow}.    
\end{split}  
\end{align}
This succeeds in flipping the pseudospin of the inner dimer, but the outer dimers are in an excited state due to the odd number
of spin flips provided by $\H_1$.
Another application of $\H_1$ on each bond will fix this, but returns the inner dimer to its initial state leading to an identity term.
On a larger cluster of dimers we run into the same problem as internally flipped pseudospins leave behind excited edge dimers.
Therefore, for half-integer spins we can only generate trivial identity terms due to a mismatch between the number of $S^xS^x$ terms needed to flip the inner spins, and
the number required return the outer dimer to is initial state
\begin{equation}
  \H_{\text{eff}} \propto \id.
\end{equation}  
In fact, we are simply seeing that the $2^{N_c}$ degeneracy in the dimerized limit cannot be lifted, as it is protected by the $N_c$ non-local SU(2) symmetries of
Eq.~(\ref{eq:nonlocalSU2}).

\subsection{Integer Spins}

In contrast to the half-integer case, there is no conflict between flipping the inner pseudospins and returning the edge dimers to their original state.
Consider a particular $S = 1$ process on three dimers for $K_0 < 0$ where we apply $S^xS^x$ on each bond $2S = 2$ times
\begin{align}
  \begin{split}
    &\ket{11}{\color{red}-}\ket{11}{\color{red}-}\ket{11} \\
    &\hspace{2.5mm}S^-\downarrow S^- \\
    &\ket{10}{\color{red}-}\ket{01}{\color{red}-}\ket{11} \\
    &\hspace{12mm}S^-\downarrow S^- \\
    &\ket{10}{\color{red}-}\ket{00}{\color{red}-}\ket{01} \\
    &\hspace{2.5mm}S^+\downarrow S^- \\
    &\ket{11}{\color{red}-}\ket{\overline{1}0}{\color{red}-}\ket{01} \\
    &\hspace{12mm}S^-\downarrow S^+ \\
    &\ket{11}{\color{red}-}\ket{\overline{1}\overline{1}}{\color{red}-}\ket{11}.
  \end{split}    
\end{align}
In this case the pseudospin of the inner dimer is flipped, {\it and} the edge dimers have returned to their initial state, resulting in a $\tau^x$ term at
order $4S = 4$ with coefficient $\sim K_1^{4S}/K_0^{4S-1}$.

Next we can consider four dimers with three $\H_1$ bonds, and two inner dimers as shown in Fig.~\ref{fig:dimers}(c).
As before the outer spin of the outer two dimers are fixed so their partners must return to their initial state.
Clearly the two subclusters of three consecutive dimers can generate a $\tau^x$ term at order $4S$ as discussed earlier.
If we completely flip the pseudospins of the inner two dimers we can generate a new term represented by $\tau^x\tau^x$.
To flip all four spins in the pair of dimers we need to apply $S^xS^x$ $2S$ times on each bond for a total of $6S$.
This gives us a $\tau^x\tau^x$ term at order $6S$ with coefficient $\sim K_1^{6S}/K_0^{6S-1}$.

The $\tau_j^x\tau_{j+1}^x$ term seen at order $6S$ involves neighbouring dimers, but in larger clusters we can generate interactions between further neighbours.
Consider five dimers with four $\H_1$ bonds and three inner dimers as shown in Fig.~\ref{fig:dimers}(d).
The three subclusters of three consecutive dimers can generate $\tau^x$ terms at order $4S$, and the two subclusters of four consecutive dimers can generate adjacent
$\tau^x\tau^x$ terms at order $6S$.
An interaction $\tau_j^x\tau_{j+2}^x$ between the non-adjacent inner dimers can be generated at order $2\cdot(4S) = 8S$ through two disjoint $\tau^x$ processes.
Finally, a three-pseudospin interaction $\tau^x\tau^x\tau^x$ appears at order $8S$ with coefficient $\sim K_1^{8S}/K_0^{8S-1}$.

Clearly this pattern will continue for arbitrarily large clusters of dimers.
Suppose we have $N_i + 2$ dimers, two outer and $N_i$ inner with $N_i + 1$ $\H_1$ bonds connecting them.
Subclusters of consecutive dimers will generate the lower order $(\tau^x)^{\otimes M}$ terms considered before for $M < N_i$.
By applying $S^xS^x$ $2S$ times on each of the $N_i+1$ bonds we can generate the highest order term $(\tau^x)^{\otimes N_i}$ term at order $2S(N_i+1)$ with coefficient
$\sim K_1^{2S(N_i+1)}/K_0^{2S(N_i+1)-1}$.

Having exhausted all the possibilities we can write down the form of the effective Hamiltonian
\begin{align}\label{eq:HeffPBC}
  \begin{split}
    \H_{\text{eff}} = &-h_{\text{eff}}^x\sum_{j}\tau_j^x \\
                      &-J_{2,1}\sum_{j}\tau_j^x\tau_{j+1}^x - J_{2,2}\sum_{j}\tau_j^x\tau_{j+2}^x - \cdots \\
                      &-J_{3,1}\sum_{j}\tau_j^x\tau_{j+1}^x\tau_{j+2}^x - J_{3,2}\sum_{j}\tau_j^x\tau_{j+1}^x\tau_{j+3}^x -\cdots,
  \end{split}    
\end{align}
which contains arbitrary strings of $\tau^x$ interactions between adjacent, and non-adjacent dimers.
The leading order contributions to the effective model parameters scale as
\begin{align}
\begin{split}  
  h_{\text{eff}}^x &\simeq f_1(S)|K_0|g^{4S}, \\
  J_{2,1} &\simeq f_{2,1}(S)|K_0|g^{6S}, \\
  J_{2,2} &\simeq f_{2,2}(S)|K_0|g^{8S}, \\
  J_{3,1} &\simeq f_{3,1}(S)|K_0|g^{8S}, \cdots
\end{split}
\end{align}
where $f_1(S),f_{2,1}(S),\cdots$ are coefficients that depend on $S$ and $g = K_1/K_0 \ll 1$.
Looking back at the representation of $W_{0/1}$ in the low energy subspace Eq.~(\ref{eq:Wlowenergy}) we see that $W_0 \mapsto \tau^x$.
So the first non-trivial term $h_{\text{eff}}^x$ selects the state with $W_0 \equiv +1$ out of the $2^{N_c}$ degenerate states, and all higher order
interactions among the $W_0$ reinforce this as the ground state at all scales.
Note that since $W_1 \mapsto \id$, its excitations are completely frozen out in the low-energy subspace.

In this discussion we have been able to identify all of the the allowed terms in $\H_{\text{eff}}$, as well as the order of perturbation theory they appear at.
However, we have not classified all the processes that give rise to these terms.
While this is an interesting combinatorial problem we will be content with explicit expressions for low $S$ and their limiting behaviour as $S \rightarrow \infty$.
For $S = 1$ the leading expressions are
\begin{align}\label{eq:S1coeffs}
\begin{split}  
  h_{\text{eff}}^x\lvert_{S = 1} &\simeq \frac{7}{192}\frac{K_1^4}{|K_0|^3} + \frac{133}{\num{6144}}\frac{K_1^6}{|K_0|^5} + \mathcal{O}\left(\frac{K_1^8}{|K_0|^7}\right),\\
  J_{2,1}\lvert_{S = 1} &\simeq \frac{449}{\num{36864}}\frac{K_1^6}{|K_0|^5} + \mathcal{O}\left(\frac{K_1^8}{|K_0|^7}\right),
\end{split}  
\end{align}
from order $4S = 4$ up to $6S = 6$ with ratio
\begin{equation}
  \frac{J_{2,1}}{h_{\text{eff}}^x}\biggr\lvert_{S = 1} \simeq \frac{449}{1344}\left(\frac{K_1}{K_0}\right)^2 \simeq \num{0.334}\ g^2.
\end{equation}
For $S = 2$ the leading expression is
\begin{equation}
  h_{\text{eff}}^x\lvert_{S = 2}\ \simeq \frac{\num{18604521}}{\num{5138022400}}\frac{K_1^8}{|K_0|^7} + \mathcal{O}\left(\frac{K_1^{10}}{|K_0|^{9}}\right),
\end{equation}
at order $4S = 8$.
We note the leading order $h_{\text{eff}}^x$ is approximately reduced by an order of magnitude
\begin{equation}
  \frac{h_{\text{eff}}^x\lvert_{S = 2}}{h_{\text{eff}}^x\lvert_{S = 1}} = \frac{\num{55813563}}{\num{561971200}}\left(\frac{K_1}{K_0}\right)^4 \simeq \num{0.099}\ g^4,
\end{equation}
from $S = 1$ to $S = 2$.

As noted earlier the interaction terms first appear at order $4S,6S,\cdots$ with a coefficient that scales as $|K_0|g^{4S},|K_0|g^{6S},\cdots$.
Letting $S \rightarrow \infty$ at some $g < 1$, these terms are pushed to higher and higher orders of perturbation theory and suppressed by a factor
$g^{2S(M+1)} \rightarrow 0$.
So in the classical limit all of the terms in $\H_{\text{eff}}$ tend to zero, and we recover the $2^{N_c}$-fold degeneracy enjoyed by the half-integer spins.
At finite $S$ the lowest order term $h_{\text{eff}}^{x} \sim |K_0| g^{4S}$ is most important, as the subleading $J_{2,1}$ interaction is suppressed by a factor of $\sim g^{2S}$.
Thus the bandwidth and the internal structure of the $2^{N_c}-1$ excited states in the low-energy subspace decreases, with $h_{\text{eff}}^x$ mainly governing their
hierarchy.

\subsubsection{Open Boundary Conditions}

The integer spin effective Hamiltonian as written in Eq.~(\ref{eq:HeffPBC}) assumes PBCs in the original model, but from our discussion
of the perturbation process it is easy to see what happens when we have OBCs.
An interaction $(\tau_x)^{\otimes M}$ among a string of $M$ dimers can be generated so long as there are an additional two dimers on either side participating.
Any dimer on the boundary consists of one spin involved in $\H_1$, and one spin that is not.
As there is no way to flip the outermost spin its partner in the dimer must always return to its initial state after any perturbative process.
This means that none of the generated interactions $(\tau_x)^{\otimes M}$ involve the edge dimer
\begin{align}\label{eq:HeffOBC}
\begin{split}  
\hspace{-5mm}  \H_{\text{eff}} = &-h_{\text{eff}}^x\sum_{j = 1}^{N_c-2}\tau_j^x \\
                    &-J_{2,1} \sum_{j = 1}^{N_c-3}\tau_j^x\tau_{j+1}^x - J_{2,2}\sum_{j = 1}^{N_c-4}\tau_j^x\tau_{j+2}^x-\cdots\\
                    &-J_{3,1} \sum_{j = 1}^{N_c-4}\tau_j^x\tau_{j+1}^x\tau_{j+2}^x - J_{3,2}\sum_{j = 1}^{N_c-5}\tau_j^x\tau_{j+1}^x\tau_{j+3}^x - \cdots,
\end{split}  
\end{align}
and there is a free pseudospin on each edge, $\tau_0^x$ and $\tau_{N_c-1}^x$.
The interactions in $\H_{\text{eff}}$ select a uniform $W_0 \equiv +1$ state for the $N_c-2$ inner dimers, but the $2^2 = 4$ states of the edge dimers are all degenerate.
We therefore predict that the anisotropic integer spin Kitaev chains support a unique ground state under PBCs, but a 4-fold degenerate state under OBCs due to free edge pseudospins.

\subsubsection{Excitations \& Thermodynamics}

Knowing the structure of the effective Hamiltonian in the anisotropic limit we can construct the excitations and low-temperature part of thermodynamic quantities.
The half-integer chains maintain the $2^{N_c}$-fold degeneracy, and thus has no low-temperature peak in the heat capacity.
However, the entropy per unit cell shows a $\log(2)$ plateau that extends down to zero temperature.

In the integer case we have seen that the initial degeneracy of $2^{N_c}$ is lifted by the perturbative terms, leaving us with a unique $W_0 \equiv +1$ ground
state and $2^{N_c-1}$ close-by states with energy $\sim h_{\text{eff}}^x$.
The $2^{N_c}-1$ low-lying excitations on top of the ground state are simply $W_0$ flips on a dimer, with an internal hierarchy determined by the
strength of the effective couplings $h_{\text{eff}}^x,J_{2,1},\cdots$ which further depend on the anisotropy and spin.
For instance, the lowest-lying excitations are $N_c$ single $W_0$ flips on each dimer that come with a cost of $2h_{\text{eff}}^x + 4 J_{2,1} + \cdots$.
These low-lying states produce a peak in the heat capacity around $k T^*_1 \simeq h_{\text{eff}}^x$, and a plateau of $\log(2)$ in the entropy per unit cell.
At low temperatures the conserved quantities are approximately their ground state values $\braket{W_{0/1}} \approx +1$.
As as we approach $T^*_1$ the value of  $\braket{W_0}$ approaches zero as we can access states with arbitrary flipped $W_0$.
On the other hand $W_1$ is frozen to $+1$ in the low-energy subspace and will maintain its value of $+1$ until temperatures on the order of $|K_0|$.

\begin{figure}
  \includegraphics[width=0.85\linewidth]{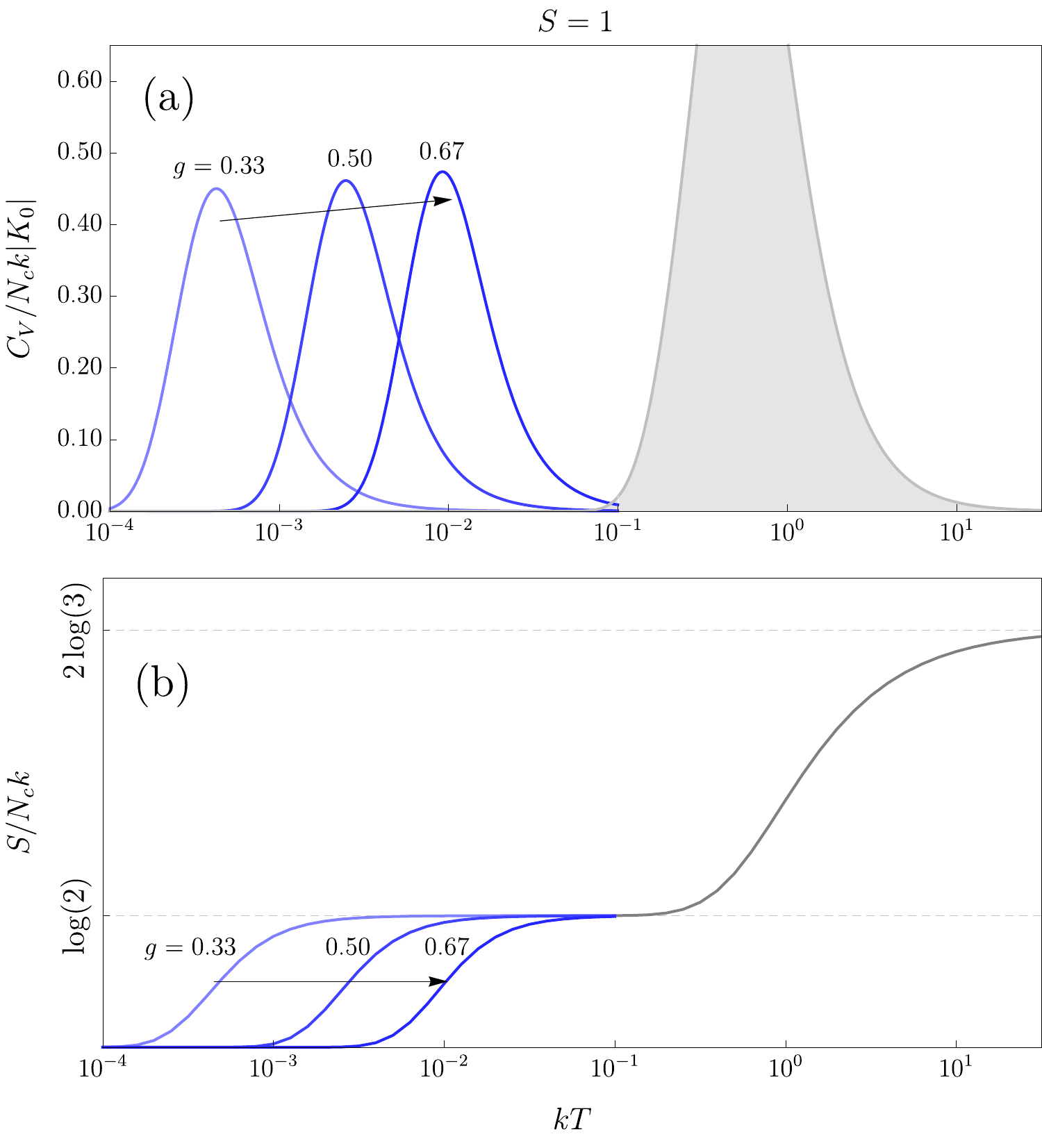}
  \caption{
    Low-temperature thermodynamics of the $S = 1$ effective Hamiltonian using the coupling constants of Eq.~(\ref{eq:S1coeffs}) up to $\mathcal{O}(g^8)$.
    (a) Heat capacity and (b) thermal entropy per unit cell of $\H_{\text{eff}}$ for $g = 0.35,0.50,0.67$ are drawn in blue. The high-energy component not captured by the effective model is
    drawn in grey.
  }
  \label{fig:s1thermo}
\end{figure}

There are two limits of integer spins we can consider.
Fixing $S = 1$ we can use the coupling constants of Eq.~(\ref{eq:S1coeffs}) to approximate $\H_{\text{eff}}$ up to $\mathcal{O}(g^8)$ including $h_{\text{eff}}^x$ and $J_{2,1}$.
Within this approximation the heat capacity and thermal entropy per unit cell are shown in Fig.~\ref{fig:s1thermo} for $g = 0.35,0.50,$ and $0.67$.
Low-temperature peaks in the heat capacity are concentrated around the dominant energy scale $h_{\text{eff}}^x$.
Increasing $g$ serves to decrease the anisotropy, and pushes this energy scale to higher temperatures. 
We have seen that all the terms in $\H_{\text{eff}}$ reinforce the unique $W_0 \equiv +1$ ground state, so we expect that the full anisotropic model shares this
property for a large part of the phase diagram.
Of course without tracking the identity terms and the energy shift of the excited states, we cannot rule out a first order transition as we approach the isotropic limit.
At the same time as the $W_0$ excitations increase in energy, the $W_1$ excitations in the high-energy subspace are decreasing as $g \rightarrow 1$.
Alternatively, approaching from the limit $g \gg 1$ the roles of $W_{0/1}$ are reversed.
We speculate that the generically the heat capacity will have a double-peak structure with the low-temperature piece representing $W_{j,0}$ excitations.
As we approach the isotropic limit we expect that a piece of the higher-temperature peak representing the $W_1$ excitations will split off and lower in energy, eventually
merging with the low-temperature peak when $g \rightarrow 1$.

\begin{figure}
  \includegraphics[width=0.85\linewidth]{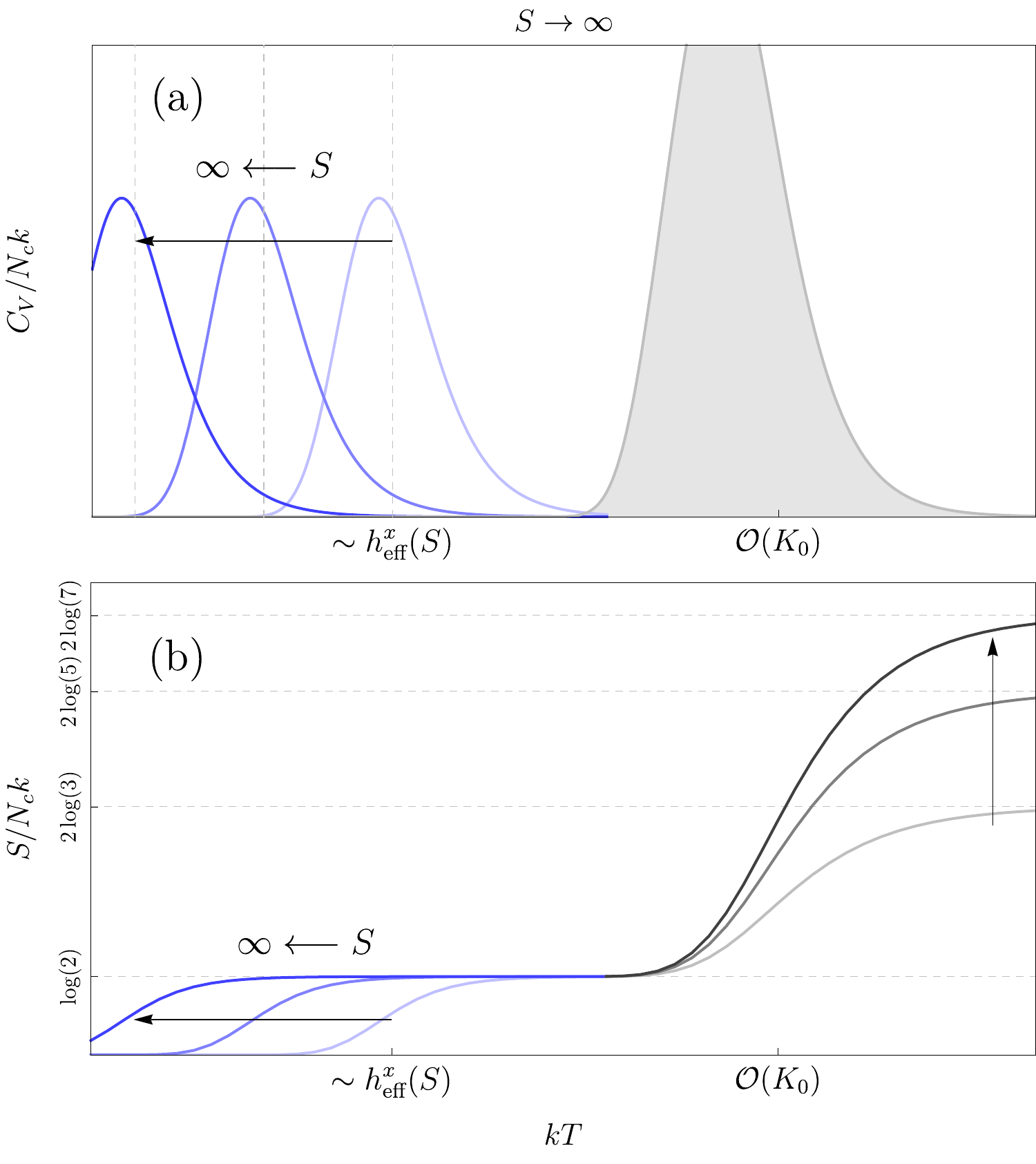}
  \caption{
    Schematic dependence of the (a) heat capacity, and (b) thermal entropy per unit cell as a function of increasing integer $S$ at a fixed anisotropy $g$.
    The dominant energy scale $h_{\text{eff}}^x(S)$ of the $W_0$ excitations rapidly decreases as a function of $S$, pushing the (a) heat capacity peak,
    and (b) the $\log(2)$ thermal entropy plateau to zero temperature, like the half-integer and classical chains.
    Again the contribution from the high-energy subspace not captured by $\H_{\text{eff}}$ is drawn in grey, and such that $S/N_c k \rightarrow 2\log(2S+1)$.
  }
  \label{fig:schematicthermo}
\end{figure}

Another limit we can consider is one of fixed anisotropy $g < 1$ and increasing integer spin $S$ as illustrated in Fig.~\ref{fig:schematicthermo}.
Initially the low-temperature peak appears at the energy scale $h_{\text{eff}}^x$, and a high-temperature peak appears around the $|K_0|$ energy scale.
As we increase $S$ the leading-order effective interaction $h_{\text{eff}}^x \propto g^{4S}$ scales to zero, as do all the higher-order terms.
This pushes the low-temperature peak in the heat capacity to lower and lower temperatures, and extends the $\log(2)$ plateau in the entropy.
Additionally, the range over which $\braket{W_0} \approx +1$ is pushed to lowest temperatures.
At large $S$ the excitations and thermodynamic quantities approach that of the half-integer and classical spin chains.

\section{Classical \& Semiclassical Limits}\label{sec:semiclassical}

The classical ground state space (CGSS) of the Kitaev chain were first elucidated by \citet{baskaran2007exact}.
Special members of the CGSS, called Cartesian states, are one for which the interactions on all the $x$- or all the $y$-bonds are fully satisfied.
Explicitly, an $x$-Cartesian state is one for which the spin configuration is
\begin{equation}\label{eq:xcartS}
  \S_{j,A} = \eta_jS\xh, \qquad \S_{j,B} = -\sgn(K_0)\eta_jS\xh,
\end{equation}
parameterized by Ising variables $\eta_j\in\{\pm 1\}$, and similarly for the $y$-Cartesian states.
If there are $N_c$ unit cells then there are $2^{N_c}$ $x$-Cartesian states, as each $\eta_j$ can be specified independently.
Without loss of generality we will assume $K_{0,1} > 0$.

In the isotropic limit the $x$- and $y$-Cartesian states have the same energy, as do any states that interpolate between them, for instance
\begin{align}\label{eq:cartparam}
\begin{split} 
  \S_{j,A} &= \cos\vp\left[+\eta_j S\xh\right] + \sin\vp\left[+\nu_{j-1}S\yh\right], \\
  \S_{j,B} &= \cos\vp\left[-\eta_jS\xh\right] + \sin\vp\left[-\nu_j S\yh\right],
\end{split}  
\end{align}
where $\vp\in[0,\tfrac{\pi}{2}]$ and the $\{\nu_j\}$ parameterize the $y$-Cartesian states.
So the classical ground state space (CGSS) is made up of an exponential number of Cartesian states, as well as an exponentially large set of continuous families that
interpolates between each $x$- and each $y$-Cartesian state.
Due to the structure of the CGSS, the phenomenon of order by singularity~\cite{khatua2019effective,khatua2021selection} becomes relevant.
The low-energy physics can be described by a particle moving on the CGSS, which has a non-manifold structure due to self-intersections at the Cartesian
points~\cite{khatua2021selection}.
Remarkably, the lowest energy states of this particle are localized near these singular points~\cite{khatua2019effective}, representing a state selection mechanism
completely distinct from the usual order by disorder (which also selects the Cartesian states~\cite{baskaran2007exact}).

In the anisotropic limit the CGSS is purely discrete, so the phenomenon of order by singularity is not relevant.
Without loss of generality we take $K_0 > K_1 > 0$ so that $0 < g = K_1/K_0 < 1$.
The $2^{N_c}$ ground states are the $x$-Cartesian states of Eq.~(\ref{eq:xcartS}) with energy $E_0 = -N_cK_0S^2$.
Averaging over the $\{\eta_j\}$ configurations, both the magnetizations $\S_{j,A/B}$ and the spin-spin correlations
\begin{equation}
  \S_{j,A/B}\cdot\S_{k,A/B} = -\bm{S}_{j,A}\cdot\S_{k,B} = S^2\eta_j\eta_k,
\end{equation}
vanish.
Our goal in this section is to find the spin wave excitations on top of the Cartesian states, their effect on the degeneracy, and their physical interpretation.

\subsection{Linear Spin Wave Theory}

Formally, we perform linear spin wave theory around each of the $2^{N_c}$ classical ground states specified by the $\Z_2$ $\{\eta_j\}$ variables in Eq.~(\ref{eq:xcartS}).
We choose the site-dependent frames
\begin{align}
\begin{split}  
  (\xh_{j,A},\yh_{j,A},\zh_{j,A}) &= (\yh,+\eta_j\zh,+\eta_j\xh), \\
  (\xh_{j,B},\yh_{j,B},\zh_{j,B}) &= (\yh,-\eta_j\zh,-\eta_j\xh),  
\end{split}  
\end{align}
so that $\zh_k$ is aligned with the local ordering direction and the linearized Holstein-Primakoff transformation~\cite{holstein1940field} becomes
\begin{align}
  \begin{split}
    \S_{k} &\sim \sqrt{\frac{S}{2}}\left[(\ad_k + \a_k)\xh_k + i(\ad_k - \a_k)\yh_k\right] + (S-\ad_k\a_k)\zh_k, \\
           &\sim \sqrt{S}(\hat{X}_k\xh_k + \hat{P}_k\yh_k) + \left(S - \frac{\hat{P}_{k}^2 + \hat{X}_{k}^2}{2}\right)\zh_k
  \end{split}    
\end{align}
with one type of boson for each sublattice $\hat{a},\hat{b}$, or equivalently in terms of canonically conjugate position- and momentum-like variables
\begin{equation}\label{eq:SHOvars}
  \hat{X}_k = \frac{1}{\sqrt{2}}(\ad_k + \a_k), \qquad \hat{P}_k = \frac{i}{\sqrt{2}}(\ad_k-\a_k).
\end{equation}
The $\hat{X}_k$ and $\hat{P}_k$ represent transverse spin fluctuations in the global $\yh$, and $\zh$ directions, respectively.

The strong $x$-bonds simply become a sum of $N_s = 2N_c$ harmonic oscillator Hamiltonians 
\begin{align}
\begin{split}
  \H_0 &\sim \frac{K_0S}{2}\sum_j\left(\hat{P}_{j,A}^2 + \hat{P}_{j,B}^2 + \hat{X}_{j,A}^2 + \hat{X}_{j,B}^2\right), \\
       &\sim K_0S\sum_j\left(\ad_j\a_j + \bd_j\b_j\right), 
\end{split}  
\end{align}
with frequency $K_0S$.
On the other hand the weaker $y$-bonds connecting the unit cells contribute a quadratic $\hat{X}_k$ potential
\begin{align}
\begin{split}
  \H_1  &\sim K_1S\sum_j\hat{X}_{j,B}\hat{X}_{j+1,A} \\
        &\sim \frac{K_1S}{2}\sum_j(\bd_j + \b_j)(\ad_{j+1} + \a_{j+1}).
\end{split}  
\end{align}
Note that even though a generic $\{\eta_j\}$ configuration does not respect translational symmetry, the resulting spin wave Hamiltonian does.
This is made possible through our choice of local frames $\{\xh_k,\yh_k\} \sim \{\yh,\pm \zh\}$; if we shifted the signs $\eta_j$ onto the $\yh$ coordinate the
signs of the $\hat{X}$ potential from $\H_1$ would have signs modulated by $\eta_j\eta_{j+1}$.
An exact solution using the real-space Bogoliubov-de Gennes Hamiltonian in that case is still possible, and yields identical results.
Transforming between the two amounts to judicious $\pi$ rotations about the local ordering direction.

In momentum space the $\hat{X}$ potential becomes
\begin{equation}
  \frac{K_0S}{2}\sum_{q}\pmat{\hat{X}_{-q,A} & \hat{X}_{-q,B}}\pmat{1 & g e^{-iqa} \\ g e^{iqa} & 1}\pmat{\hat{X}_{q,A} \\ \hat{X}_{q,B}},
\end{equation}
which can be diagonalized through the transformation
\begin{equation}
  \hat{X}_{q,\pm} = \frac{1}{\sqrt{2}}\left(\hat{X}_{q,A} \pm e^{-iqa}\hat{X}_{q,B}\right),
\end{equation}
into
\begin{equation}\label{eq:XpotentialSWT}
  \frac{K_0S}{2}\sum_{q}\left[(1 + g)\hat{X}_{-q,+}\hat{X}_{q,+} + (1-g)\hat{X}_{-q,-}\hat{X}_{q,-}\right].
\end{equation}
The kinetic energy term is unchanged by the corresponding transformation of $\hat{P}$, so we see that the spin wave excitations have uniform
frequencies $\omega_{\pm} = K_0S\sqrt{1 \pm g}$.
With OBCs there will be additional mode(s) with frequency $\omega_0 = K_0S$ between these two branches associated with the edge spin(s)
not participating in $y$-bonds.
Note that since this result is independent of the $\{\eta_j\}$ configuration, the spin wave fluctuations do not lift the classical $2^{N_c}$ degeneracy. 
In the isotropic limit the lower linear spin wave theory band vanishes $\omega_{q,-} \rightarrow 0$, signalling additional members of the CGSS Eq.~(\ref{eq:cartparam})
that connect different Cartesian states.

An explicit construction of the transformation shows us that the excitations are completely localized on the $y$-bonds, and can be understood through 
certain deformations of the spins.
Consider two unit cells $j-1$ and $j$ with local ordering along the $\eta_{j-1}\xh$ and $\eta_j\xh$ directions.
The peripheral spins $\S_{j-1,A} = \eta_{j-1}S\xh$ and $\S_{j,B} = \eta_jS\xh$ are unaffected, but we deform the adjacent inner spins forming the $y$-bond slightly via
\begin{align}
  \begin{split}
    \S_{j-1,B} &= S\left[-\eta_{j-1}\cos\vp\ \xh \pm \sin\vp\ \yh\right], \\
    \S_{j,A} &= S\left[\eta_{j}\cos\vp\ \xh \pm \nu \sin\vp\ \yh\right],     
  \end{split}    
\end{align}
where $\vp \ll 1$ and $\nu \in \{\pm 1\}$.
Note that $\nu = \pm 1$ represents a deformation where the $\yh$ components of the neighbouring spins have the same/opposite signs.
Due to the spin length constraint, the energy on the two affected $x$-bonds is raised by
\begin{equation*}
  \Delta E_x = 2\left(1 - \cos\vp \right)K_0S^2 \simeq K_0S^2\vp^2,
\end{equation*}
while the energy on the central $y$-bond becomes
\begin{equation*}
  \Delta E_y = \nu K_1S^2\sin^2\vp \simeq K_0S^2\nu g \vp^2,
\end{equation*}
for a total of
\begin{equation}
  \Delta E \simeq K_0S^2(1 + \nu g)\vp^2 \sim (1 + \nu g)\vp^2.
\end{equation}
This is a harmonic potential of frequency $\omega \sim \sqrt{1 \pm g}$.
Since $K_1 > 0$ the deformations with the $\yh$ components of the spins aligned ($\nu = +1$) costs more than those anti-aligned ($\nu = -1$).
If we instead deformed the spins in the $\zh$ direction, the only energy cost is on the two affected $x$-bonds producing a potential of
\begin{equation}
  \Delta E \simeq K_0S^2\vp^2.
\end{equation}
These two wells produce the potential in $\hat{X}$ and kinetic energy of $\hat{P}$, respectively.
As we approach the isotropic limit $g \rightarrow 1^-$ the kinetic energy terms stay finite, while one of the $\hat{X}$ potentials vanishes.
The soft deformation actually corresponds to new members of the CGSS that interpolates between the $x$- and $y$-Cartesian states.
In the language of~\citet{rau2018pseudo} these are local type I zero modes, which are called improper or spurious in the language of~\citet{colpa1986zeromodeI,colpa1986zeromodeII}.

Finally, the predicted correction to the ordered moment from linear spin wave theory is
\begin{equation}
  \braket{S_{j,A/B}^x} = \pm\eta_jS\left[1 - \frac{m(g)}{S}\right],
\end{equation}
where
\begin{equation}
m(g) = \frac{1}{8}\left[\frac{(1-\sqrt{1+g})^2}{\sqrt{1+g}} + \frac{(1-\sqrt{1-g})^2}{\sqrt{1-g}}\right].
\end{equation}
As we approach the isotropic limit $g \rightarrow 1^-$ the contribution from the soft mode diverges since those $\yh$ fluctuations are less and less bound by
the restoring potential.

It is important to distinguish the excitations found within spin wave theory from those found at finite $S$ in the perturbation theory.
The effective spin-$S$ model describes the $\Z_2$ degrees of freedom within a unit cell that correspond to the $2^{N_c}$ classical $x$-Cartesian ground states.
A unique $W_0 \equiv 1$ ground state is selected for integer spins, and the excitations are $W_0 = -1$ defects with dispersion governed by the interactions $\{J_{2,1},\cdots \}$.
The bandwidth of these $2^{N_c}-1$ excitations is also set by the interactions $\{h_{\text{eff}}^x,J_{2,1},\cdots \}$, which are rapidly decaying functions of increasing $S$ at fixed anisotropy.
As $S \rightarrow \infty$ both the gap and bandwidth vanish, leaving us with the $2^{N_c}$ degeneracy enjoyed by half-integer spins and the classical $x$-Cartesian ground states.

On the other hand, the linear spin wave theory describes small fluctuations around {\it each} of the classically degenerate $2^{N_c}$ $x$-Cartesian states.
These excitations are found to be completely localized on the $y$-bonds and involve spin deformations not captured in the low-energy subspace of the spin-$S$ effective model.
With finite anisotropy the spin wave excitations are gapped since the $y$-Cartesian states are higher in energy.
As the isotropic limit is approached the energy of the $y$-Cartesian states approaches that of the $x$-Cartesian states, and the potential barrier between them softens.
Consequently, the lower spin wave excitation energy approaches zero, reflecting the fact that the $y$-Cartesian states become part of the classical ground state manifold.

\section{Summary \& Conclusions}

In summary we have studied the general anisotropic spin-$S$ Kitaev chain in the quantum, classical, and semiclassical limits.
Expanding around the extensive $2^{N_c}$ ground state degeneracy of the dimerized limit, we determine the structure of the effective Hamiltonian at finite
anisotropy to all orders in perturbation theory.
For half-integer spins we find a trivial effective Hamiltonian that does not lift the $2^{N_c}$ degeneracy, reflecting their protection from a set of $N_c$ non-local SU(2) symmetries.
With integer spins there is no such symmetry protecting the degeneracy, and the effective Hamiltonian contains a large number of non-trivial terms.
For integer $S$ the important degrees of freedom in the low-energy subspace are found to be the $\Z_2$ $W_{j,0}$ degrees of freedom on the strong $x$-bonds, with $W_{j,1}$ on the weaker $y$-bonds
being frozen out.
Due to an effective magnetic field of the $W_{j,0}$ at order $4S$, a unique $W_{0} \equiv +1$ state is selected.
Higher-order ferromagnetic interactions among the $W_{j,0}$ are generated at higher orders, starting at $6S$, which emphatically select the flux-free ground state and imparts dispersion onto the excitations.
Since all the effective interactions reinforce the unique flux-free ground state we expect that a large part of the phase diagram with anisotropy has this property, potentially including the isotropic limit.

This effective model leads to a natural description of the thermodynamic features discovered recently in the $S = 1$ chain~\cite{luo2021unusual}.
A unique $W \equiv +1$ ground state is found across all anisotropies, and the $N_c$ low-lying excitations correspond to flipping $W$ on single bonds.
The low-temperature peak in the heat capacity is naturally explained by the $2^{N_c}-1$ $W_{0}$ flux excitations on the strong bonds, and leads to a $\log(2)$ plateau in the entropy per unit cell
at intermediate temperatures.
We find that the temperature scale of the low-temperature peak is set by the leading-order $h_{\text{eff}}^x$ of our effective Hamiltonian, and as we approach the isotropic limit this energy scale increases.
At fixed anisotropy increasing the integer spin length $S$ serves to decrease the magnitude of all the effective interactions.
This has the effect of reducing the temperature scale of the low-temperature peak in the heat capacity, the bandwidth of the low-energy excitations, and the height of this peak.
In the limit $S \rightarrow \infty$ we regain the $2^{N_c}$ degeneracy of the half-integer spins, and the classical $x$-Cartesian states.

We have also studied the anisotropic Kitaev spin chain the classical and semi-classical limits with linear spin wave theory.
There are $2^{N_c}$ degenerate $x$-Cartesian states at the classical level, which we study spin wave fluctuations around.
The spin wave excitations are found to be completely localized on the $y$-bonds, and reflect spin deformations not captured in the low-energy subspace of our earlier effective model.
As these frequencies are the same for each Cartesian state, spin wave fluctuations do not break the $2^{N_c}$ degeneracy.
Approaching the isotropic limit causes one of these frequencies to approach zero energy due to the fact that the $y$-Cartesian states have joined the classical ground state manifold.
Based on these findings we can speculate on the isotropic limit of the integer spin-$S$ Kitaev chains.
We expect a unique ground state with $W \equiv +1$ to persist into the isotropic limit, with a gap that rapidly decreases as a function of $S$.
With anisotropy there is a double-peak structure in the heat capacity, and the lower-temperature peak corresponds to excitations of $W$ on the strong bonds.
As we approach the isotropic limit the energy of the corresponding excitations on the weak bonds comes down in energy, and approaches that of the strong bonds.
So we expect at some small anisotropy a subset of the states comprising the upper peak in the heat capacity will come down in energy and merge with the low-temperature peak in the isotropic limit.
Further studies of the thermodynamic properties of higher integer-spin Kitaev chains are needed to test this conjecture.

\begin{acknowledgments}
  We would like to thank Q. Luo and P.P. Stavropoulos for useful discussion.
  This work was supported by the Natural Sciences and Engineering Research Council of Canada (NSERC) Discovery Grant No. 2016-06089,
  the Center for Quantum Materials at the University of Toronto, the Canadian Institute for Advanced Research (CIFAR), and the
  Canada Research Chairs Program.
  Computations were performed on the Niagara supercomputer at the SciNet HPC Consortium.
  SciNet is funded by: the Canada Foundation for Innovation under the auspices of Compute Canada; the Government of Ontario; Ontario Research Fund - Research Excellence;
  and the University of Toronto.
\end{acknowledgments}

\appendix

\section{Details of the Spin Wave Theory}\label{appendix:SWTdetails}

In the basis of Holstein-Primakoff bosons we can write the spin wave Hamiltonian in Bogoliubov-de Gennes form
\begin{equation}
  \H = \frac{K_0S}{2}\sum_q\hat{\Psi}_q^{\dag}\D_q\hat{\Psi}_q
\end{equation}  
where $\hat{\Psi}_q^T = \pmat{\a_q & \b_q & \ad_{-q} & \bd_{-q}}$ and
\begin{equation}
  \D_q = \id_4 + \frac{1}{2}g\pmat{0 & e^{-iqa} & 0 & e^{-iqa} \\ e^{+iqa} & 0 & e^{+iqa} & 0 \\ 0 & e^{-iqa} & 0 & e^{-iqa} \\ e^{+iqa} & 0 & e^{+iqa} & 0}.
\end{equation}
The algorithm of~\citet{colpa1978diagonalization} can always be used to diagonalize this numerically, but here we will construct the paraunitary transformation
explicitly through a sequence of transformations through harmonic oscillator fields as in Eq.~(\ref{eq:SHOvars}).
Our first transformation is to the oscillator fields on each site
\begin{equation}
  \pmat{\hat{P}_{q,A} \\ \hat{P}_{q,B} \\ \hat{X}_{q,A} \\ \hat{X}_{q,B}} =
  \underbrace{\frac{1}{\sqrt{2}}\pmat{-i & 0 & +i & 0 \\ 0 & -i & 0 & +i \\ 1 & 0 & 1 & 0 \\ 0 & 1 & 0 & 1}}_{\Tds_1}\pmat{\a_q \\ \b_q \\ \ad_{-q} \\ \bd_{-q}},
\end{equation}
which separates the Hamiltonian into kinetic and potential energy sectors.
Normal modes are found by rotating the position and momentum operators to diagonalize the potential energy
\begin{equation}
  \pmat{\hat{P}_{q,+} \\ \hat{P}_{q,-} \\ \hat{X}_{q,+} \\ \hat{X}_{q,-}} =
  \underbrace{\frac{1}{\sqrt{2}}\pmat{1 & +e^{-iqa} & 0 & 0 \\ 1 & -e^{-iqa} & 0 & 0 \\ 0 & 0 & 1 & +e^{-iqa} \\ 0 & 0 & 1 & -e^{-iqa}}}_{\Tds_2}
  \pmat{\hat{P}_{q,A} \\ \hat{P}_{q,B} \\ \hat{X}_{q,A} \\ \hat{X}_{q,B}},
\end{equation}
At this point we recognize the (dimensionless) frequencies of $\D_q$ as $\varpi_{\pm} = \sqrt{1 \pm g}$ from the transformed potential in Eq.~(\ref{eq:XpotentialSWT}).
The frequencies of the spin wave Hamiltonian are simply $\omega_{\pm} = K_0S\varpi_{\pm} = K_0S\sqrt{1 \pm g}$.
It is then straightforward to find the transformation into bosonic operators $\gamma_{q,\pm}$
\begin{equation}
  \hspace{-5mm}\pmat{\g_{q,+} \\ \g_{q,-} \\ \gd_{-q,+} \\ \gd_{-q,-}} =
  \underbrace{\frac{1}{\sqrt{2}}\pmat{+i\varpi_+^{-1/2} & 0 & \varpi_+^{1/2} & 0 \\ 0 & +i\varpi_-^{-1/2} & 0 & \varpi_-^{1/2} \\
    -i\varpi_+^{-1/2} & 0 & \varpi_+^{1/2} & 0 \\ 0 & -i\varpi_-^{-1/2} & 0 & \varpi_-^{1/2}}}_{\Tds_3}
  \pmat{\hat{P}_{q,+} \\ \hat{P}_{q,-} \\ \hat{X}_{q,+} \\ \hat{X}_{q,-}}.
\end{equation}
Therefore we have found a diagonalizing transformation
\begin{equation}
  \hat{\Gamma}_q = \pmat{\g_{q,+} \\ \g_{q,-} \\ \gd_{-q,+} \\ \gd_{-q,-}} = \Tds_q\pmat{\a_q \\ \b_q \\ \ad_{-q} \\ \bd_{-q}} = \Tds_q\hat{A}_q,
\end{equation}
where $\Tds_q = \Tds_3\Tds_2\Tds_1$.
We can verify that $\Tds_q$ satisfies the paraunitary condition $\Tds_q\J\Tds_q^{\dag} = \J$ where
\begin{equation*}
  \J = \pmat{+\id_2 & 0 \\ 0 & -\id_2},
\end{equation*}
and therefore preserves the bosonic commutation relations.
The explicit form of $\Tds_q$ is
\begin{equation}
  \Tds_q =\pmat{
    +T_1 & +e^{-iqa}T_1 & -T_3 & -e^{-iqa}T_3 \\
    +T_2 & -e^{-iqa}T_2 & -T_4 & +e^{-iqa}T_4 \\
    -T_3 & -e^{-iqa}T_3 & +T_1 & +e^{-iqa}T_1 \\
    -T_4 & +e^{-iqa}T_4 & +T_2 & -e^{-iqa}T_2},
\end{equation}
where
\begin{align*}
  \begin{split}
    T_1 &= \frac{1 + \varpi_+}{2\sqrt{2}\sqrt{\varpi_+}} = \frac{1 + \sqrt{1 + g}}{2\sqrt{2}(1 + g)^{1/4}}, \\
    T_2 &= \frac{1 + \varpi_-}{2\sqrt{2}\sqrt{\varpi_-}} = \frac{1 + \sqrt{1 - g}}{2\sqrt{2}(1 - g)^{1/4}}, \\
    T_3 &= \frac{1 - \varpi_+}{2\sqrt{2}\sqrt{\varpi_+}} = \frac{1 - \sqrt{1 + g}}{2\sqrt{2}(1 + g)^{1/4}}, \\
    T_4 &= \frac{1 - \varpi_-}{2\sqrt{2}\sqrt{\varpi_-}} = \frac{1 - \sqrt{1 - g}}{2\sqrt{2}(1 - g)^{1/4}}.    
  \end{split}    
\end{align*}
To translate this back into real-space we simply need to inverse Fourier transform $\hat{\Gamma}_q$
\begin{equation}
  \hat{\Gamma}_j = \pmat{\g_{j,+} \\ \g_{j,-} \\ \gd_{j,+} \\ \gd_{j,-}} = \frac{1}{\sqrt{N_c}}\sum_qe^{iqx_j}\Tds_q\hat{A}_q.
\end{equation}
The $\a_{q}$ and $\ad_{-q}$ are not met with any $q$ dependence in $\Tds_q$, so the inverse Fourier transform simply transforms them to $\a_j$ and $\ad_{j}$, respectively.
On the other hand both $\b_q$ and $\bd_{-q}$ are imparted a phase factor $e^{-iqa}$ from $\Tds_{q}$ which serves to shift their real-space counterparts one unit cell left
as $\b_{j-1}$ and $\bd_{j-1}$, respectively.
This is because the excitations are localized on the $y$-bonds which involve $\S_{j,A}$ and $\S_{j-1,B}$.
Therefore, the real-space transformation is simply
\begin{equation}
  \pmat{\g_{j,+} \\ \g_{j,-} \\ \gd_{j,+} \\ \gd_{j,-}} =
  \underbrace{
    \pmat{+T_1 & +T_1 & -T_3 & -T_3 \\ +T_2 & -T_2 & -T_4 & +T_4 \\ -T_3 & -T_3 & +T_1 & +T_1 \\ -T_4 & +T_4 & +T_2 & -T_2}}_{\Tds_j}
  \pmat{\a_j \\ \b_{j-1} \\ \ad_j \\ \bd_{j-1}},
\end{equation}
where the subscript in $\Tds_j$ now refers to the unit cell and not the transformations making up $\Tds_q$ from earlier.
The spin-wave corrected ground state is a vacuum for $\gamma_{j,\pm}$ involving lowering of the spins on the $A$ and $B$ sites that form the $y$-bonds.
Using the paraunitary nature of $\Tds_j$, we can calculate the inverse transformation via $\mathbb{S}_j = (\Tds_j)^{-1} = \J\Tds_j^{\dag}\J$
\begin{equation}
  \pmat{\a_j \\ \b_{j-1} \\ \ad_j \\ \bd_{j-1}} =
  \underbrace{
    \pmat{+T_1 & +T_2 & +T_3 & +T_4 \\ +T_1 & -T_2 & +T_3 & -T_4 \\ +T_3 & +T_4 & +T_1 & +T_2 \\ +T_3 & -T_4 & +T_1 & -T_2}}_{\mathbb{S}_j}\pmat{\g_{j,+} \\ \g_{j,-} \\ \gd_{j,+} \\ \gd_{j,-}}.
\end{equation}
Time-dependence of the operators is easily accounted for since  $\g_{j,\pm}(t) = e^{-i\omega_{\pm}t}\g_{j,\pm}$ and $\gd_{j,\pm}(t) = e^{+i\omega_{\pm}t}\gd_{j,\pm}$.

Before calculating the physical spin quantities, we first discuss the isotropic limit $g \rightarrow 1^{-}$.
In this limit we cannot use the transformation directly as $\Tds$ is singular.
However, from Eq.~(\ref{eq:XpotentialSWT}) we can see that the restoring potential for $\hat{X}_{q,-}$ vanishes in the isotropic limit, while the kinetic energy for
$\hat{P}_{q,-}$ remains finite.
In the language of~\citet{rau2018pseudo} this corresponds to a type I zero mode at each wavevector, where the canonically conjugate partner of $\hat{X}_{q,-}$ costs
finite energy.
This can also be verified through the criteria of~\citet{colpa1986zeromodeI,colpa1986zeromodeII}.
Eigenvalues of $\D_q$ at $g = 1$ are solutions of
\begin{equation*}
  \det\left(\varpi\id - \D_q\right) = \varpi(\varpi - 1)^2(\varpi-2) = 0,
\end{equation*}
while the paravalues of $\D_q$ at $g = 1$ are solutions of
\begin{equation*}
  \det\left(\varpi\id - \J\D_q\right) = \varpi^2(\varpi^2 - 2) = 0
\end{equation*}
So we see that there is a single zero eigenvalue and a double zero paravalue of $\D_q$ at each wavevector.
In the language of~\citet{colpa1986zeromodeII} this means that the zero mode is improper, or spurious.
Physically this represents the fact that these spin deformations in the $\yh$ directions become members of the ground state manifold as discussed in the main text.

Having constructed the transformation explicitly, we can now calculate quantities involving the spins.
Going back to the linearized Holstein-Primakoff transformation, we easily find $\braket{S_{j,A/B}^{y/z}} = 0$ and the spin wave correction to the $x$ moments
from their classical values
\begin{align}
  \begin{split}
    \left\lvert\frac{\eta_jS - \braket{S_{j,A}^x}}{S}\right\lvert &= \frac{\braket{\ad_j\a_j}}{S} = \frac{m(g)}{S} \\
    \left\lvert\frac{\overline{\eta_j}S - \braket{S_{j,B}^x}}{S}\right\lvert &= \frac{\braket{\bd_j\b_j}}{S} = \frac{m(g)}{S}
  \end{split}    
\end{align}
where
\begin{align}
\begin{split}  
  m(g) &= T_3^2 + T_4^2 \\
         &= \frac{1}{8}\left[\frac{(1 - \sqrt{1 + g})^2}{\sqrt{1 + g}} + \frac{(1 - \sqrt{1 - g})^2}{\sqrt{1 - g}}\right].
\end{split}  
\end{align}
As expected, the correction vanishes $m(g) \rightarrow 0$ as we approach the dimerized limit $g \rightarrow 0^+$.
Note that the second term in $m(g)$ diverges as $g \rightarrow 1^-$, so the moment will eventually flip (effectively changing $\eta_j \mapsto \overline{\eta_j}$) before
becoming unphysical at some critical $g$.
In reality the spin wave theory will break down before this extreme limit, since the decreasing potential barrier between the Cartesian states allows for
tunnelling between them.
This aspect is captured by the perturbation theory at finite $S$.

\bibliography{references}\label{sec:bib}

\end{document}